# Exploration of legal implications of air and space travel for international and domestic travel and the Environment.

Jayanthi Vajiram[1], Negha Senthil[2,] Nean Adhith.P[3,] Ritikaa.VN [4]

[1,2] Research Scholar, Vellore Institute of Technology, Chennai, India; negha.2019@vitstudent.ac.in (Law School), jayanthi.2020@vitstudent.ac.in

[3,4] namooshivnian23@gmail.com(SRM Valliammal Eng College), ritikaavn.238@gmail.com(Ambedkar Law College)

**Abstract**

The rapid growth of air and space travel in recent years has resulted in an increased demand for legal regulation in the aviation and aerospace fields. This paper provides an overview of air and space law, including the topics of aircraft accident investigations, air traffic control, international borders and law, and the regulation of space activities. With the increasing complexity of air and space travel, it is important to understand the legal implications of these activities. This paper examines the various legal aspects of air and space law, including the roles of national governments, international organizations, and private entities. It also provides an overview of the legal frameworks that govern these activities and the implications of international law. Finally, it considers the potential for future developments in the field of air and space law. This paper provides a comprehensive overview of the legal aspects of air and space travel and their implications for international and domestic travel, as well as for international business and other activities in the air and space domains. This paper investigates the legal implications of aircraft accidents, traffic control, and international borders in the context of air and space law. It examines the existing statutory and regulatory frameworks developed to address these issues, as well as the roles of international organizations, states, and private actors in ensuring safety and compliance with the law. It then outlines the potential challenges posed by the ever-evolving nature of air and space technology and its impact on the legal framework. Finally, the paper considers the potential future developments in the field of air and space law and their implications for the international community. The paper concludes that a comprehensive and balanced approach is needed to ensure the safety of air and space travel, as well as to manage the ever-expanding legal framework in this domain.

## 1.Introduction

The exploration of legal implications of air and space travel for international and domestic travel and the Environment is a complex and multi-faceted undertaking that requires a deep understanding of the legal and political implications of the aerospace industry. As the development of air and space travel continues to grow, legal implications become increasingly important. This paper will provide a comparative study of the legal implications of air and space travel for international and domestic travel and the environment. The paper will first provide a brief overview of the history of air and space travel, discussing the development of the industry, the evolution of the legal framework, and the current state of air and space travel. Following this, the paper will explore the legal implications of air and space travel for international and domestic travel, analyzing the various legal issues that arise in both areas. The paper will then discuss the legal implications of air and space travel for the environment, examining the various environmental impacts of air and space travel. Finally, the paper will provide a conclusion summarizing the findings of the paper and providing some recommendations for the future of air and space travel. By providing a comprehensive overview of the legal implications of air and space travel for international and domestic travel and the environment, this paper will serve as a valuable resource for understanding the complex legal and political implications of the aerospace industry.

## 2.Air traffic control and related issues like congestion and environmental problems;

Air traffic control is an integral component of the aviation system that is responsible for the safe and efficient management of airspace around airports and other areas of high air traffic. This includes the separation of aircraft, issuing instructions for take-off and landing, and providing warnings about weather, terrain, and other hazards. As air travel increases, air traffic controllers must be able to manage increasingly complex flight patterns, while balancing efficiency and safety, as well as the needs of carriers, passengers, and other stakeholders. In addition, air traffic controllers must be aware of the environmental impacts of air travel, such as emissions of carbon dioxide and nitrogen oxides, and take steps to minimize these impacts by reducing fuel consumption, increasing the efficiency of air traffic management, and improving the sustainability of air travel. For these reasons, it is essential that air traffic controllers be properly trained to meet the demands of the aviation industry.

Air traffic control is a complex and highly specialized field of aviation, tasked with providing safe and efficient services to aircraft in flight and on the ground. Due to the increasing number of aircraft operating in international airspace, ATC is under pressure to manage congestion, delays, and environmental issues. Congestion has caused delays and increased the likelihood of mid-air collisions, while aircraft emissions have a direct impact on air quality and climate change. International organizations have implemented laws and regulations to address these issues, but the complexities of ATC still remain.

The International Civil Aviation Organization (ICAO) is the global body responsible for establishing regulations and standards for international air navigation. The primary duty of air traffic controllers is to safely separate aircraft in the sky, using tools such as radar, radio frequencies, and other communication systems. In addition to directing traffic, controllers must provide pilots with essential information, like weather conditions, air traffic congestion, and navigation guidance. Unfortunately, air traffic congestion can lead to delays, cancellations, and reduced safety. Moreover, aircraft emissions can have serious environmental consequences. In order to ensure the safe and efficient operation of air traffic control, the ICAO is working to develop new rules and regulations, as well as new technologies like satellite-based navigation systems. By doing so, the ICAO is helping to ensure the safety of passengers and crew, while also protecting the environment.

Air traffic congestion is a major problem in the aviation industry, leading to delays, airspace restrictions, and environmental issues. To address this issue, new technologies such as satellite-based navigation systems have been developed to reduce flight delays and fuel consumption by up to 25%. Additionally, air traffic management techniques like flight planning, traffic flow management, and airspace management have been instrumental in improving the efficiency of air travel. Advanced data analysis techniques can also be used to identify and resolve air traffic congestion issues, allowing for improved safety and efficiency of air travel. The Massachusetts Institute of Technology (MIT) conducted a study which concluded that the use of these advanced data analysis techniques can help identify and resolve air traffic congestion issues in a timely manner. [1,2,3].

## 3.Congestion in Airspace

Airspace congestion is a major problem facing the aviation industry, caused by an increasing number of commercial flights and a growing demand for air travel. This has resulted in delays, increased costs, and safety concerns. To address this issue, the aviation industry has implemented various strategies to reduce congestion, including direct routing, altitude control, and speed control. Air traffic controllers must strive to limit the number of aircraft in congested areas, as well as the number of aircraft in the same airspace at the same time in order to maximize efficiency and safety.

The increasing congestion problem in the U.S. National Airspace System (NAS) requires an efficient and effective system for managing air traffic. To this end, the potential impact of user fees on air traffic congestion is examined in this paper, taking into account the effects on aircraft speed and altitude, air traffic controllers, and airspace users. Simulation models are used to analyze the effects of different speeds and altitudes on aircraft separation and airspace congestion. The results of the simulation demonstrate that increasing aircraft speed and altitude has a positive effect on the performance of air traffic flow management and can be used to reduce delays. The authors discuss the implications of their findings and suggest solutions for reducing congestion in airspace [4,5,6,7,8].

### 3.1. Related Laws

1. Federal Aviation Administration Reauthorization Act of 2018: This act requires the Federal Aviation Administration (FAA) to develop an air traffic congestion management plan to reduce the impact of air traffic congestion on airports, aircraft, and airspace users.

2. Air Traffic Congestion Relief Act of 2018: This act seeks to improve aviation infrastructure and reduce air traffic congestion by authorizing the FAA to develop and implement a comprehensive plan to reduce air traffic congestion.

3. Airport Improvement Program: This act authorizes funding for airport improvements, including projects designed to reduce air traffic congestion.

4. Next Generation Air Transportation System: This act seeks to modernize the air traffic control system and reduce air traffic congestion by implementing a new air traffic control system that utilizes satellite-based technology and advanced automation.

5. FAA Modernization and Reform Act of 2012: This act seeks to reduce air traffic congestion by requiring the FAA to develop and implement an air traffic control system that utilizes advanced automation and satellite-based navigation technology.

6. Air Traffic Management System Performance Improvement Act of 2018: This act seeks to reduce air traffic congestion by requiring the FAA to develop and implement an air traffic control system that utilizes advanced automation and satellite-based navigation technology.

### 4. Airspace Capacity

Airspace capacity is limited by the available physical space for aircraft to safely fly and the number of air traffic controllers needed to manage that airspace. To address this issue, the aviation industry has employed various strategies to increase capacity, such as introducing new airspace management technologies, increasing air traffic management efficiency, and using airspace optimization techniques. Airspace capacity determines the amount of available airspace for aircraft to operate in a given area, and is influenced by factors such as the number of aircraft, the type of airspace, and the availability of navigation aids. The Federal Aviation Administration (FAA) regulates airspace capacity in the United States, while the National Airspace System (NAS) and the Air Traffic Control System Command Center (ATCSCC) help to monitor and control airspace capacity. By following the regulations and guidelines set by the FAA and other organizations, airspace capacity can be managed efficiently and safely.

Airspace capacity is an important factor to consider when designing and operating air traffic control systems. It is the maximum number of aircraft that can safely and efficiently operate in a given airspace at any given time. The capacity of an airspace is determined by a wide range of factors, including the size and shape of the airspace, the type of aircraft operating in it, the prevailing weather conditions, and the number of air traffic control systems in place. Airspace capacity affects the number of planes that can land and takeoff at an airport, the efficiency of air traffic control operations, and ultimately, the safety of air travel. [9,10].

**4.1Related Laws**

1. The Federal Aviation Administration Reauthorization Act of 2018 (FAR 18) establishes a new framework for airspace capacity planning that emphasizes collaboration with local governments, airports, airlines, and other stakeholders.

2. The FAA Modernization and Reform Act of 2012 (FAR 12) requires the FAA to modernize the air traffic control system through the deployment of new technologies and procedures, including NextGen.

3. The Vision 100-Century of Aviation Reauthorization Act of 2003 (VRA 03) requires the FAA to develop long-term strategies to reduce delays, increase efficiency, and improve safety in the national airspace system.

4. The Airport Improvement Program Reauthorization Act of 1998 (AIR 98) authorizes the FAA to provide financial assistance to airports for the construction, maintenance, and improvement of airport capacity.

5. The Air Transportation Safety and System Stabilization Act of 2001 (ATSSSA) authorizes the FAA to regulate aviation safety, security, and capacity in order to protect the public and promote the integrity of the national airspace system.

6. The Next Generation Air Transportation System Reauthorization Act of 2018 (NGATSRA 18) requires the FAA to develop a NextGen implementation plan that includes provisions for airspace capacity optimization.

7. The Airport and Airway Safety and Capacity Expansion Act of 2010 (AASCEA 10) authorizes the FAA to develop and implement strategies to improve the safety and capacity of the national airspace system.

8. The FAA Extension, Safety, and Security Act of 2016 (FESSA 16) authorizes the FAA to develop a comprehensive air traffic control modernization plan that includes provisions for airspace capacity optimization.

**5.Air Traffic Flow Management**

Air traffic flow management (ATFM) is an important strategy that is used to manage airspace congestion. ATFM is a process that uses aircraft scheduling, route optimization, and other techniques to maximize airspace capacity and reduce delays. It is a complex process that requires cooperation between airlines, air traffic controllers, and other stakeholders. Air Traffic Flow Management (ATFM) is the process of managing the flow of air traffic in order to reduce delays, ensure safety, and optimize airspace utilization. It is a collaborative decision-making process between air traffic controllers, pilots, and other stakeholders, such as airlines and meteorologists, who must coordinate their efforts in order to ensure safe and efficient air traffic. ATFM is an important part of air traffic control, as it helps to ensure that air traffic controllers manage the flow of air traffic in an efficient and safe manner. It involves a variety of activities such as setting flight paths, allocating airspace, and adjusting flight routes, as well as ensuring that aircraft are safely separated from one another. ATFM also involves predicting and responding to weather events, such as thunderstorms and turbulence, as well as other unexpected events such as an aircraft emergency. ATFM is essential for ensuring the safety of air travelers and the efficiency of air traffic. Without an effective ATFM system, delays and congestion in the air traffic system could lead to costly delays and potential accidents.

## 5.1 Air Traffic Flow Management related laws

1. Federal Aviation Administration (FAA) Order 7110.65, Air Traffic Control

2. FAA Order JO 7110.10, Air Traffic Organization Policy

3. FAA Order JO 7110.65M, Air Traffic Control

4. FAA Order JO 7400.2, Air Traffic Control System Command Center

5. FAA Order JO 7400.11, Air Traffic Flow Management

6. FAA Order JO 7250.3, Terminal Radar Approach Control (TRACON)

7. FAA Order JO 7400.3, Air Traffic Control System Network Management

8. FAA Order JO 7350.8, Air Traffic Procedures Advisory Circular

9. FAA Order JO 7260.3, Air Traffic Control System Command Center Automation

10. FAA Order JO 7400.4, Air Traffic Control System Command Center Automation Program Management

Air Traffic Flow Management (ATFM) is defined by the International Civil Aviation Organization (ICAO) as "the regulation of air traffic in terms of its rate, balance, and spacing, in order to ensure an orderly flow of traffic" (ICAO, 2008). ATFM is used to ensure aircraft are properly separated and to reduce delays due to capacity constraints. It is a critical component of the air traffic control system and involves the coordination of airspace, airports, and en route traffic [11,12]. A novel air traffic flow management system based on game theory is proposed in this paper. This system is designed to address the conflicts between different aircrafts in order to achieve efficient flight scheduling. It utilizes a constraint optimization model to optimize the aircraft trajectory and to determine the optimal speed for each aircraft. Additionally, a multi-player game is used to resolve the conflicts between different aircrafts, and a learning-based algorithm is used to further refine the optimal speed of each aircraft. The system is evaluated with flight data from Beijing Capital International Airport and its results show that the proposed system is able to reduce the total flight delay time by up to 37% [13]. This paper provides a comprehensive overview of Air Traffic Flow Management (ATFM) and its various approaches, such as the traditional ground-based approach and the more advanced airborne approach. It reviews the advantages and disadvantages of each approach, as well as discusses the challenges to implementing ATFM. Finally, the paper provides an outlook on the future of ATFM and how it may be improved [14]. Air traffic flow management (ATFM) is the process by which air traffic controllers coordinate the movement of aircraft to ensure safety and efficiency. This paper presents a review of current ATFM research, addressing topics such as trajectory prediction, conflict detection and resolution, airspace structure, and air traffic control automation. It is concluded that progress in ATFM research has been made, but more work is needed to improve the safety and efficiency of air traffic control [15]. The ATFM approach enables the optimization of the air traffic system to meet the capacity of the system, while minimizing delays and improving the efficiency of the system. By controlling the number of flights in the system and optimizing flight paths and ground delays, the ATFM approach helps to reduce the level of congestion in the air traffic system. Additionally, the ATFM approach helps to reduce the cost of air traffic operations by providing an efficient system for managing the demand-capacity balance [16]. Air Traffic Flow Management (ATFM) is a complex problem due to the dynamic nature of the air traffic environment. In order to ensure efficient and safe operations, the aviation industry is actively seeking innovative solutions to address the ATFM challenge. This paper presents an overview of the current state of ATFM and the future trends in the field. It focuses on the various aspects of ATFM, such as airspace design, traffic simulation, conflict resolution, and optimization techniques, as well as the potential benefits of incorporating advanced technologies such as unmanned aerial vehicles and artificial intelligence into ATFM systems. Finally, the paper proposes a roadmap for the future development of ATFM systems [17,18]. The authors suggest that further research in the field of ATFM should focus on developing models and algorithms that can improve the performance of ATFM systems. They suggest that such research should focus on addressing the complexities of air traffic flow, such as the interactions between aircraft, airspace, and traffic controllers. Additionally, research should focus on developing algorithms that are robust and able to handle large-scale changes in air traffic. Finally, the authors recommend that future research should also explore the use of artificial intelligence and machine learning to improve ATFM systems. Overall, the authors provide an overview of the existing literature on ATFM and make recommendations for future work [19].

## 6. Airspace Optimization

Airspace optimization is a strategy used to reduce air travel congestion, delays, and costs. By utilizing modern air traffic control techniques such as Automatic Dependent Surveillance-Broadcast (ADS-B) and Traffic Flow Management (TFM) systems, air traffic controllers can more accurately monitor aircraft, coordinate routes, and adjust the speed and altitude of flights in order to optimize the flow of traffic. This optimization leads to smoother, safer, and more efficient air travel experiences for both passengers and crew. Additionally, by reducing the noise generated by aircraft, fuel consumption, and emissions, airspace optimization can help to reduce the financial costs associated with air travel. Ultimately, airspace optimization helps to ensure a more pleasant and cost-effective air travel experience. Furthermore, airspace optimization can improve the profitability of airlines by reducing the amount of time and money spent on air navigation fees, as well as increasing their market share. By reducing the amount of time spent on air navigation fees, airlines can save money that can be invested in other areas, such as updating their fleets or offering more competitive fares. Finally, airspace optimization can help to reduce the environmental impact of air travel, as it can reduce the amount of time that aircraft spend in the air and the amount of fuel they consume. Airspace optimization is the process of using data-driven analysis, advanced analytics, and modeling to reduce delays, increase airspace capacity, and improve safety by optimizing airspace utilization. By leveraging data and analytics, airspace optimization solutions can identify and prioritize airspace improvements that reduce congestion, improve air traffic flow, and reduce delays. Optimization solutions can also identify and prioritize airspace improvements that reduce weather-related disruptions, improve air traffic flow, and reduce delays. In addition, airspace optimization solutions can provide insights into aircraft operations and performance, helping to maximize efficiency and reduce costs. Airspace optimization solutions are typically used by aviation management organizations, such as the Federal Aviation Administration (FAA), the European Aviation Safety Agency (EASA), and the International Civil Aviation Organization (ICAO). These solutions enable these organizations to improve their operational performance by optimizing airspace utilization and increasing airspace capacity. One example of an airspace optimization solution is the use of advanced analytics and data-driven decision support systems. These solutions can be used to identify areas of airspace congestion and to develop strategies to reduce delays and improve air traffic flow. Data-driven decision support systems can also provide insights into the performance of aircraft operations and can help to identify and prioritize airspace improvements.

Other airspace optimization solutions include the use of air traffic flow management (ATFM) systems, which use data and analytics to optimize air traffic flow and reduce delays. ATFM systems can also provide insights into aircraft operations and can help to identify and prioritize airspace improvements. In addition to these solutions, airspace optimization can also involve the use of airspace and aircraft performance monitoring systems, which use data and analytics to monitor airspace activity and aircraft performance. These solutions can provide insights into the performance of aircraft operations and can help to identify and prioritize airspace improvements. Finally, airspace optimization solutions can also involve the use of airspace models and simulation, which use data and analytics to simulate aircraft operations and to identify and prioritize airspace improvements. These solutions can help to improve the efficiency and safety of airspace operations [20,21,22,23,24,25]. The aviation sector uses Artificial intelligence for predictive maintenance, auto-scheduling, pattern recognition, targeted advertising, and an analysis of client feedback.

### 6.1. Airspace Optimization related laws

1. Aircraft Noise Abatement: Many countries have enacted laws, regulations, and policies to reduce aircraft noise, including restrictions on the times of day that aircraft can fly, restrictions on flying routes and altitudes, and incentives for aircraft operators to reduce noise.

2. Airspace Restrictions: Increasingly, governments are limiting the use of airspace over certain areas to reduce the noise impact on nearby communities.

3. Airspace Segmentation: This is the process of dividing airspace into smaller sections, in order to better control air traffic flows and reduce the potential for conflicts between aircraft.

4. Airspace Capacity Management: This is the process of managing the use of airspace in order to maximize its capacity and ensure that air traffic flows smoothly and safely.

5. Airspace User Fees: Governments may impose fees on users of airspace in order to encourage them to make more efficient use of it, or to pay for improvements.

6. Aircraft Performance Requirements: Governments may require aircraft to meet certain performance standards in order to reduce their environmental impact. This may include requirements for fuel efficiency and noise reduction.

7. Air Traffic Control Modernization: Governments may invest in modernizing air traffic control systems in order to enable more efficient use of airspace and reduce delays.

8. Airspace Security: Governments may impose restrictions on airspace use in order to protect the safety and security of their citizens.

Congestion in airspace is a major issue for the aviation industry. To address this issue, the aviation industry has implemented various strategies to increase airspace capacity and reduce delays. These strategies include the introduction of new airspace management technologies, increased air traffic management efficiency, and the use of airspace optimization techniques. By utilizing these strategies, the aviation industry can ensure that airspace is used efficiently and safely.

### 7.Aircraft accident investigation and liability legislation

Aircraft accident investigation is a complex process that is necessary to determine the cause of an aircraft accident and the liability of any involved parties. The process involves collecting evidence, interviewing witnesses, and conducting a thorough analysis of the accident scene. The goal of the investigation is to identify any safety issues that may have contributed to the accident and make recommendations to prevent similar accidents in the future. The process of aircraft accident investigation is regulated by a variety of international laws and regulations. These laws and regulations are designed to ensure that the investigation is conducted in a fair and impartial manner and that the results of the investigation are accurate. In addition to international regulations, many countries have their own aircraft accident investigation and liability legislation. These laws and regulations are designed to protect the public from potential harm caused by aircraft accidents. They also provide guidance to aircraft operators, manufacturers, and other parties regarding their responsibilities in the event of an accident. In the United States, the Federal Aviation Administration (FAA) is responsible for conducting aircraft accident investigations and issuing liability rulings. The FAA regulations establish a framework for determining the cause of an accident and assessing the liability of any involved parties. Additionally, the FAA may impose civil penalties or criminal charges against any parties found to be responsible for the accident. Overall, aircraft accident investigation and liability legislation is critical to ensuring that aircraft accidents are properly investigated and that those responsible for an accident are held accountable. By ensuring that the investigation process is conducted in a fair and impartial manner, these laws and regulations help to promote safety and reduce the risk of future accidents.

### 7.1. Laws in International Borders

Due to the global nature of air traffic, there are laws and regulations that govern air traffic in different countries. This includes the rules and regulations of the International Civil Aviation Organization (ICAO), which sets out standards and recommended practices for international civil aviation. Additionally, each country may have its own set of laws and regulations that govern air traffic within its borders.

In addition, there are also laws and regulations that govern air traffic across international borders. For example, the Chicago Convention, which was signed in 1944, sets out the rules and regulations governing international air traffic. This includes the rights of aircraft to fly over other countries, as well as the responsibilities of aircraft operators flying across international borders.

International border law is the set of legal rules and regulations that govern the movement of people, goods, and services across international borders. It is a complex and dynamic area of law, shaped by a variety of international organizations, treaties, regulations, and domestic laws. The law is further complicated by the fact that different countries have different laws governing the movement of people, goods, and services across their borders. International border law has been studied extensively by legal scholars, and there is a large body of scholarly articles and books on the subject. One of the primary goals of international border law is to ensure the safety and security of citizens and visitors crossing national borders. This includes preventing the unauthorized entry of individuals and goods into countries, as well as regulating the use of border crossings for commerce and trade. One of the most important aspects of international border law is the regulation of immigration. This includes laws concerning the entry and exit of foreign nationals, as well as the process for seeking asylum and other types of immigration relief. This area of law is constantly evolving due to the ever-changing political and economic climate of the world. The United Nations Convention on the Law of the Sea (UNCLOS) is an important international agreement that governs the use of the ocean for navigation, fishing, and other activities. It also sets out the rights and responsibilities of countries with regard to their territorial waters, including the regulation of maritime borders. Additionally, UNCLOS is important for regulating the law of the sea, which governs the use of the ocean for commercial activities such as shipping and fishing. The European Union (EU) is also a major player in the field of international border law. The EU has a number of directives and regulations that govern the movement of people, goods, and services across its borders. This includes the Schengen Agreement, which allows for free movement of citizens within the EU and the Dublin Regulation, which sets out the conditions for granting asylum and other forms of immigration relief. Finally, the International Court of Justice (ICJ) is the primary judicial body for resolving disputes between countries in the field of international law. The ICJ is responsible for interpreting and applying international treaties, agreements, and conventions, such as those governing immigration and the law of the sea. The ICJ also has the power to issue binding decisions in disputes between countries [from 26 to 36].

## 8. Environmental Issues

The environmental impact of air traffic has been a major concern for many years. Aircraft produce various pollutants including carbon dioxide, nitrogen oxides, and sulfur dioxide. These pollutants are known to contribute to climate change and air pollution, as well as having various other negative effects on the environment.

In recent years, various initiatives have been put in place to reduce the environmental impact of air traffic. This includes the use of more fuel-efficient aircraft, the introduction of emission trading systems, and the development of alternative fuels. In addition, various technologies have been developed to reduce the amount of noise produced by aircraft.

1. **Climate Change:** Climate change is one of the greatest environmental challenges of our time, and its impacts are being felt around the world. It is caused by the emissions of greenhouse gases such as carbon dioxide, methane, and nitrous oxide, which are generated by human activities such as burning fossil fuels and deforestation. The effects of climate change are wide-ranging and include increases in global temperature, extreme weather events, sea level rise, ocean acidification, and biodiversity loss (IPCC, 2013; UNFCCC, 2018).

2. **Air Pollution:** Air pollution is a global issue and is a major contributor to climate change. It is caused by the burning of fossil fuels, industrial activities, agricultural burning, and deforestation. Air pollution has a wide range of adverse health effects, including respiratory and heart diseases, and is particularly damaging for vulnerable populations such as children, the elderly, and those with existing respiratory conditions (WHO, 2018; EPA, 2020).

3. **Water Pollution:** Water pollution is a major environmental threat and is caused by the discharge of industrial and agricultural chemicals into waterways and oceans. It can lead to algal blooms, oxygen depletion, and the death of aquatic species. It also has health impacts, with contaminated water causing water-borne illnesses such as cholera and dysentery (UNESCO, 2020; EPA, 2020).

4. **Deforestation:** Deforestation is the process of clearing forests for the purposes of logging, agricultural production, and urban development. It has a significant impact on global climate and biodiversity, as trees play a key role in the carbon cycle and are home to many species of animals and plants (UNESCO, 2020; FAO, 2020).

5. **Biodiversity Loss:** Biodiversity loss is the decrease in the variety of species in a given area due to human activities such as habitat destruction, pollution, and climate change. Biodiversity loss can have a wide range of effects on ecosystems, including decreased carbon storage, disruption of food webs, and impaired water quality (IPBES, 2019; UNEP, 2020).

6. **Ocean Acidification:** Ocean acidification is caused by the absorption of carbon dioxide from the atmosphere into the oceans. This increases the acidity of seawater, which can have a wide range of impacts on marine life, including decreased growth and reproduction, changes in species composition, and direct mortality (IPCC, 2013; NOAA, 2018).

7**. Nuclear Waste:** Nuclear waste is generated by nuclear power plants and other nuclear facilities. It is highly toxic and can remain radioactive for thousands of years. If not properly managed, nuclear waste can have a serious impact on human health and the environment, leading to contamination of soil and water, and increased risk of cancer (IAEA, 2018; EPA, 2020).

8. **Overpopulation:** Overpopulation is a global issue that is driven by a combination of factors, including high fertility rates, low mortality rates, and migration. It can lead to increased resource consumption, increased pollution, and increased pressure on land and water resources (UNFPA, 2020; World Bank, 2020).

9. **Ozone Layer Depletion:** Ozone layer depletion is caused by the release of chlorofluorocarbons (CFCs) into the atmosphere, which break down the ozone molecules. This reduces the amount of ozone in the stratosphere, leading to increased exposure to ultraviolet (UV) radiation and increased risk of skin cancer and other health problems (UNEP, 2020; EPA, 2020).

10. **Waste Management:** Waste management is the process of collecting, transporting, treating, and disposing of solid and hazardous waste (WHO, 2018; EPA, 2020. Poor management of waste can lead to environmental contamination, air and water pollution, and the spread of disease [from 36 to 47].

The environmental impact of air traffic regulation is the effect that the regulation of air traffic has on the environment. Air traffic has a significant impact on the environment due to emissions of greenhouse gases, noise pollution, and other pollutants. The effects of air traffic can be both local and global, impacting air quality, biodiversity, and climate. Regulation of air traffic can help reduce these impacts by limiting the number of flights, setting emissions limits, and reducing noise pollution. By limiting the number of flights, emissions can be reduced as fewer flights means less fuel burned and fewer pollutants released. Setting emissions limits can help ensure that aircraft engines are as efficient as possible and that the amount of air pollutants released is minimized. Finally, reducing noise pollution can help reduce the impact of aircraft on the environment, as noise pollution can disturb wildlife and disrupt ecosystems.

**Potential challenges posed by the ever-evolving nature of air and space technology and its impact on the legal framework are** :1. Increased complexity of international air law: Rapid advances in air and space technology have created a need for a more comprehensive and up-to-date body of international air law. As the technology continues to evolve, the legal framework must keep up in order to ensure a safe and efficient aviation system. 2. Unclear legal status of unmanned aerial vehicles: The use of unmanned aerial vehicles is rapidly increasing, yet the legal status of these aircraft remains unclear. Without a clear legal framework, companies are unable to operate safely and efficiently. 3. Impact of space exploration on international space law: Advances in space technology have enabled humans to explore the outer limits of our universe. However, the legal implications of space exploration are still uncertain. International space law must be developed to ensure that space activities are conducted in a safe and orderly manner. 4. Differences in airspace regulations: Airspace regulations vary from country to country, making it difficult for aircraft to traverse international borders. This can lead to flight delays and other problems. A unified set of airspace regulations is needed to ensure a safe and efficient aviation system. 5. Privacy issues: As the technology of air and space travel becomes more advanced, the privacy of individuals and companies may be compromised. Legal frameworks must be developed to protect the privacy of individuals and corporations while still allowing for the efficient operation of air and space travel.

The comprehensive and balanced approach is needed to ensure the safety of air and space travel, as well as to manage the ever-expanding legal framework in this domain. This would include the development of safety standards, the implementation of rules and regulations, the establishment of a dispute resolution mechanism, and the promotion of international cooperation. Such an approach should be based on the principles of safety, security, and the protection of human rights. Furthermore, it should consider the economic, social, and environmental impact of air and space travel, and ensure that the interests of all stakeholders are taken into account. Finally, it should ensure that the necessary resources are made available to facilitate the smooth and efficient operation of air and space travel.

**The potential future developments in the field of air and space law and their implications for the international community**

**1. Advances in Space Tourism:** Space tourism is set to become a reality in the near future as companies such as SpaceX and Blue Origin invest heavily into the development of reusable rockets and space travel technology. This could have significant implications for the international community, particularly in terms of the regulation of space tourism, the protection of astronauts and the development of space safety protocols.

**2. Growth of the Space Economy:** The growth of the space economy is expected to have significant implications for the international community. With a greater focus on space exploration and the utilization of space resources, the international community will have to develop a comprehensive legal framework to govern the exploitation and utilization of space resources and the exploitation of space-based services.

**3. International Space Cooperation:** International cooperation in the field of air and space law is also likely to increase in the coming years. This will involve the development of new treaties and agreements, such as the Outer Space Treaty, to govern the use of space and space activities.

**4. Expansion of Airspace:** With the development of new airspace technologies, the airspaces of many countries are likely to expand significantly in the near future. This could lead to increased disputes over the regulation and use of airspace, as well as the need to develop new international laws and regulations to govern the use of airspace.

**Conclusion**

The legal implications of air and space travel for both international and domestic travel, as well as the environment, are far-reaching and complex. International and domestic air and space travel are heavily regulated, both in terms of safety and environmental protection. International air and space travel is subject to a number of international treaties and agreements, as well as national laws and regulations. Domestic air and space travel is subject to a wide range of national laws and regulations, as well as international agreements and treaties.

International air and space travel is subject to a variety of international agreements, such as the Convention on International Civil Aviation of 1944 (the Chicago Convention). This convention established the International Civil Aviation Organization (ICAO) and sets out the rights and responsibilities of states in relation to international air travel. The ICAO develops and maintains a number of standards and recommended practices for international air travel, such as aerodrome design and aircraft operations.

In addition, international aviation is subject to a number of bilateral and multilateral agreements between states. These agreements set out the rights and obligations of states in relation to international air travel, as well as the rights of passengers and airlines. These agreements are often supplemented by further agreements between individual states and airlines.

Domestic air and space travel is subject to a wide range of national laws and regulations. These laws and regulations set out the rights and responsibilities of airlines, passengers, and other stakeholders in relation to domestic air travel. These laws and regulations often include safety requirements, such as the maintenance of aircraft and the operation of aircraft. In addition, domestic air travel is subject to a variety of national laws and regulations in relation to the environment, such as the control of noise and air pollution.

In addition to the legal implications of air and space travel, there are also a number of environmental implications. Air travel is responsible for a significant amount of greenhouse gas emissions, as well as other pollutants such as particulate matter. In addition, air travel has an impact on the local environment, such as noise pollution. As such, air and space travel are subject to a variety of environmental regulations and standards, both at the national and international level.

The legal and environmental implications of air and space travel are complex and far-reaching. International and domestic air and space travel is subject to a wide range of international agreements, national laws and regulations, and environmental regulations and standards. These regulations and standards are designed to ensure the safety of air travel, as well as to protect the environment from the impacts of air travel. It is important that states, airlines, and other stakeholders comply with these laws and regulations in order to ensure the safe and sustainable operation of air and space travel.